\documentclass[useAMS,usenatbib]{mn2e}
\usepackage{graphicx}

\usepackage{times}

\newcommand{\kms}{\hbox{${\rm km\,s}^{-1}$}}
\newcommand{\zem}{\hbox{$z_{\rm em}$}}
\newcommand{\zab}{\hbox{$z_{\rm abs}$}}
\newcommand{\lya}{\hbox{{\rm Ly}$\alpha$}}
\newcommand{\lyb}{\hbox{{\rm Ly}$\beta$}}
\newcommand{\snmin}{\hbox{${\rm S/N}_{\rm min}$}}
\newcommand{\gama}{\hbox{$\gamma_\alpha$}}
\newcommand{\gamr}{\hbox{$\gamma_{r^\prime}$}}
\newcommand{\gami}{\hbox{$\gamma_{i^\prime}$}}
\newcommand{\gamz}{\hbox{$\gamma_{z^\prime}$}}
\newcommand{\EBV}{\hbox{$E(B\!-\!V)$}}
\newcommand{\EWr}{\hbox{$W_{\rm r}$}}

\defcitealias{PeiY_91a}{PFB91}
\defcitealias{ProchaskaJ_04a}{PHF04}
\defcitealias{MenardB_03a}{MP03}

\def\bsp_small{\vspace{0.5cm}\small\noindent This paper has been typeset
from a \TeX/\LaTeX\ file prepared by the author.\normalsize}

\title[DLA reddening and lensing of SDSS QSOs]{Dust-reddening and
gravitational lensing of SDSS QSOs due to foreground damped
Lyman-$\bmath{\alpha}$ systems}

\author[M. T. Murphy, J. Liske]{
  M. T. Murphy$^{1}$\thanks{E-mail: mim@ast.cam.ac.uk (MTM); jliske@eso.org (JL)},
  J. Liske$^{2}$\footnotemark[1]\\
  $^{1}$Institute of Astronomy, University of Cambridge, Madingley Road,
  Cambridge, CB3 0HA, UK\\
  $^{2}$European Southern Observatory, Karl-Schwarzschild-Str.~2, 85748
  Garching, Germany}

\begin{document}

\date{Accepted ---. Received ---; in original form ---}

\pagerange{\pageref{firstpage}--\pageref{lastpage}} \pubyear{2004}

\maketitle

\label{firstpage}

\begin{abstract}
We use Sloan Digital Sky Survey Data Release 2 QSO spectra to constrain the
dust-reddening caused by intervening damped \lya\ systems (DLAs). Comparing
the spectral index distribution of a 70 sight-line DLA sample with that of
a large control sample reveals no evidence for dust-reddening at
$z\!\sim\!3$. Our limit on the shift in spectral index,
$\left|\Delta\alpha\right|\!<\!0.19$ ($3\,\sigma$), corresponds to a limit
on the colour excess due to SMC-like dust-reddening, $\EBV\!<\!0.02{\rm
\,mag}$ ($3\,\sigma$). This is inconsistent with the early studies of Fall,
Pei and collaborators who used the small QSO and DLA samples available at
the time. Comparison of the DLA and control magnitude distributions also
reveals $\ga\!2\,\sigma$ evidence for an excess of bright and/or a
deficit of faint QSOs with foreground DLAs. Higher equivalent width DLAs
give a stronger signal. We interpret this as the signature of gravitational
magnification due to the intervening DLAs.
\end{abstract}

\begin{keywords}
dust, extinction -- galaxies: high redshift -- intergalactic medium --
galaxies: ISM -- quasars: absorption lines
\end{keywords}

\section{Introduction}\label{sec:intro}

Dust, and its relationship with the gas phase, are key ingredients in any
recipe for galaxy formation and evolution. Understanding the role of dust
in the damped Lyman-$\alpha$ systems (DLAs) seen in QSO spectra is
particularly important since DLAs are thought to comprise a significant
fraction of the high redshift gas available for star-formation
\citep[e.g.][]{LanzettaK_91a}. Information about dust in DLAs comes
predominantly from the relative depletion of refractory (e.g.~Fe) and
non-refractory (e.g.~Zn) elements onto dust grains
\citep[e.g.][]{PettiniM_97a}. This measures the amount of dust in DLAs but
does little to constrain the dust's composition or grain size.

DLA dust-reddening and extinction are potentially acute problems for
flux-limited optical QSO surveys
\citep[e.g.][]{OstrikerJ_84a}. \citet{FallS_89b} examined this
observationally by comparing the spectral indices of QSO spectra with and
without foreground DLAs. They later detected \citep*{FallS_89a} and
confirmed \citep*[][hereafter \citetalias{PeiY_91a}]{PeiY_91a} a
significant difference between the spectral index distributions for the DLA
and control samples, concluding that up to 70\,per cent of bright QSOs will
be missed by flux-limited surveys \citep{FallS_93a}. A recent
radio-selected QSO survey for DLAs \citep{EllisonS_01c} finds that optical
surveys underestimate the DLA number density per unit redshift by at most a
factor of two.

The DLA and control samples of \citetalias{PeiY_91a} are quite small,
comprising 26 and 40 sight-lines respectively. The Sloan Digital Sky Survey
\citep[SDSS;][]{StoughtonC_02a} includes a large, homogeneous QSO sample
with accurate spectrophotometric calibration and spectral resolution high
enough ($R\!\approx\!1800$) to reliably detect DLAs
\citep[e.g.][]{ProchaskaJ_04a}. The SDSS therefore provides a powerful
probe of DLA dust-reddening.

Another intriguing result is the recent $\approx\!4\,\sigma$ detection by
\citet{MenardB_03a} \citepalias[hereafter][]{MenardB_03a} of gravitational
lensing (GL) caused by strong intervening Mg{\sc \,ii} absorbers in the 2dF
QSO redshift survey. By comparing the magnitude distributions of the Mg{\sc
\,ii} sample and a large control sample, they found an excess (deficit) of
bright (faint) QSOs with absorbers and demonstrated that this was
consistent with a GL interpretation. It is important to confirm and explore
the GL produced by DLAs since, in principle, it provides a probe of the
dark matter distribution in distant halos. Also, if the GL magnification of
the DLA sample is large, DLA dust-reddened QSOs will be more detectable
than those without, biasing any detection of dust-reddening.

In this paper, we select DLAs and, importantly, a large control sample of
QSOs from the SDSS Data Release 2 \citep[DR2;][]{AbazajianK_04a} to
constrain DLA dust-reddening (Section \ref{sec:dust}). We also tentatively
confirm the GL effect and discuss the potential bias on reddening (Section
\ref{sec:grav}). We present only the main observational results here,
leaving most details to a later paper.

\section{Selecting DLAs from the SDSS DR2}\label{sec:DLAs}

Spectra for all objects classified as QSOs in the SDSS DR2 with emission
redshifts $\zem\!>\!2.4$ were visually inspected and the small number
($<\!1$\,per cent) which were clearly not QSOs at the SDSS-assigned
emission redshift were rejected. We included those QSOs not listed as
`primary' targets in the SDSS, i.e.~those which were not selected using the
colour-space techniques of \citet{RichardsG_02a}. Other SDSS sources were
selected from the FIRST and ROSAT surveys or could have been initially
identified as stars or galaxies before spectroscopic follow-up. Our sample
is therefore largely, though not strictly, homogeneously selected. The
results are robust against this small inhomogeneity (Sections
\ref{ssec:dbias} \& \ref{ssec:gbias}).

We search for DLAs between the \lya\ and \lyb\ emission lines and, to avoid
DLAs intrinsic to the QSOs and moderately broad absorption lines (BALs), we
ignore the regions $<\!10000\,\kms$ below \lya\ and above \lyb. A continuum
is formed by iteratively fitting a third-order polynomial to overlapping
$20000$-\kms\ spectral chunks, rejecting pixels $>\!2\,\sigma$ below and
$>\!5\,\sigma$ above the fit at each iteration until no more points are
rejected ($\sigma$ is the SDSS error array). The continuum chunks are
combined by weighting each from zero at the edges to unity at the
centre. Finally, the continuum is smoothed over 25 pixels
($\sim\!1700\,\kms$). This procedure yields reliable continua in most
cases. However, 3 likely DLAs were not selected due to poor fits near the
continuum edges.

\begin{figure}
\centerline{\includegraphics[height=80mm,angle=270]{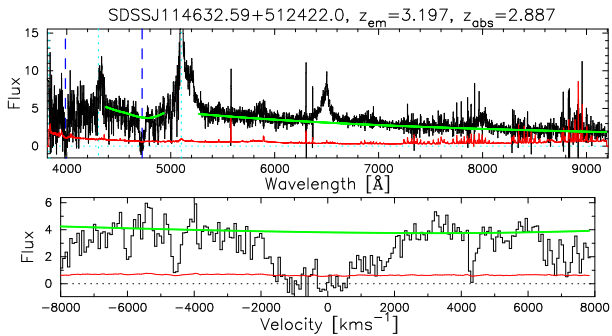}}
\caption{Top panel: Typical QSO with polynomial fit between the \lyb\ and
\lya\ emission lines and power-law fit red-wards of \lya. Dashed vertical
lines mark the \lya\ and \lyb\ absorption lines. Bottom panel: \lya\
absorption line. The lower solid line in both panels shows the SDSS
$1\,\sigma$ error array.}
\label{fig:example}
\end{figure}

Candidate DLAs are identified as absorption features with rest-frame
equivalent width $\EWr({\rm Ly}\alpha)\!\geq\!10{\rm \,\AA}$ over a
rest-frame $\Delta\lambda_{\rm r}\!=\!15{\rm \,\AA}$ window. Visual
inspection of each DLA candidate is used to reject cases where no clear DLA
profile is observed. This was the case for $\sim\!50$\,per cent of
candidates and was more prominent at $\zem\!>\!3$ where the \lya\ forest is
thicker. \citet[][hereafter \citetalias{ProchaskaJ_04a}]{ProchaskaJ_04a}
advocate a DLA search strategy where no continuum is required and this may
prove to be a more efficient future method. Nevertheless, our strategy will
select the strongest DLAs which are arguably (see below) ideal for our
study of dust-reddening and GL. Fig.~\ref{fig:example} shows an example
DLA.

We reject spectra where the median signal-to-noise per pixel --
(continuum)$_i$/($1\,\sigma$ error)$_i$ for pixel $i$ over the
$\Delta\lambda_{\rm r}\!=\!15{\rm \,\AA}$ window -- drops below a threshold
value of $\snmin\!=\!3$ anywhere along the fitted continuum. Below this the
algorithmic and visual assessment is unreliable. This is an important
selection criteria for the GL study in Section \ref{sec:grav} and we
discuss it further there. We also reject QSOs with such severe BALs that
DLA detection is unreliable, particularly those similar to LoBAL and
FeLoBALs \citep[e.g.][]{ReichardT_03a}. However, we kept QSOs with moderate
BALs (e.g.~HiBALs) since one can detect, and confidently not detect, DLAs
in these cases. This is discussed further in Sections \ref{ssec:dbias} \&
\ref{ssec:gbias}.

\begin{figure}
\centerline{\includegraphics[height=78mm,angle=270]{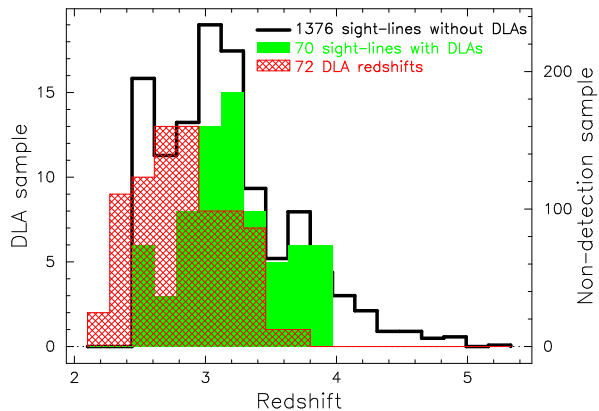}}
\caption{\zem\ distribution of QSOs with and without foreground DLAs (solid
  and hollow histograms). There are 1162 non-detections over the DLA
  sample's \zem\ range. The hatched histogram shows the \zab\
  distribution.}
\label{fig:zdist}
\end{figure}

Fig.~\ref{fig:zdist} shows the \zem\ and \zab\ distributions for QSOs where
a DLA is and is not detected. We find 72 DLAs along 70 distinct QSO
sight-lines. Although the DLA and non-detection \zem\ distributions are
similar for $2.4\!<\!\zem\!<\!4.0$, there is a high-\zem\ tail where no
DLAs are detected. This is primarily because our DLA search is insensitive
at high-\zem\ where the \lya\ forest is very thick. The total redshift path
available for DLA detection is $\Delta z\!\approx\!572$ so, using an
incidence of DLAs per unit redshift of 0.25 for a mean $z_{\rm
abs}\!=\!2.8$ \citepalias[e.g.][]{ProchaskaJ_04a}, we should have found
$\approx\!143$ DLAs. As expected, comparison with the DR1 DLA catalogue of
\citetalias{ProchaskaJ_04a} shows this 50\,per cent incompleteness to be
confined to DLAs with low neutral hydrogen column densities, $N($H{\sc
\,i}$)\!<\!10^{21}{\rm \,cm}^{-2}$. We discuss the influence this has on
our results in Sections \ref{ssec:dbias} \& \ref{ssec:gbias}.  Table
\ref{tab:cat} lists the relevant properties of the DLA and control samples.

\begin{table}
\begin{center}
\vspace{-1mm}
\caption{Catalogue of DR2 QSO and DLA properties. The J2000 name, emission
  redshift, Galactic extinction corrected $i^\prime$-band PSF magnitude,
  spectral index and \snmin\ are shown for each QSO and the absorption
  redshift and measured rest-frame \lya\ equivalent width is given for each
  candidate DLA. Here we show only a small sample from the full table which
  is available in the electronic edition of this paper and from
  http://www.ast.cam.ac.uk/$\sim$mim/pub.html. In the electronic version we
  give full name designations, $r^\prime$- and $z^\prime$-band magnitudes,
  statistical uncertainties in $\alpha$ and $W_{\rm r}$ and a series of
  flags indicating whether DLA candidates are considered genuine from
  visual inspection and whether the QSO is a `primary' target, whether it
  is included in the DR1 catalogue of \citet{SchneiderD_03a} and whether it
  contains BALs.}
\vspace{-4mm}
\label{tab:cat}
\begin{tabular}{lcccccc}\hline
\multicolumn{1}{c}{SDSSJ}&\multicolumn{1}{c}{\hspace{-1mm}\zem}&\multicolumn{1}{c}{\hspace{-1mm}$i^\prime$}&\multicolumn{1}{c}{\hspace{-1mm}$\alpha$}&\multicolumn{1}{c}{\hspace{-1mm}\snmin}&\multicolumn{1}{c}{\hspace{-1mm}\zab}&\multicolumn{1}{c}{\hspace{-1mm}\EWr}\\\hline
001240$+$135236 & \hspace{-1mm}3.187 & \hspace{-1mm}19.47 & \hspace{-1mm}$-0.55$ & \hspace{-1mm}3.02 & \hspace{-1mm}--    & \hspace{-1mm} --  \\
001255$-$091425 & \hspace{-1mm}3.004 & \hspace{-1mm}19.62 & \hspace{-1mm}$-0.07$ & \hspace{-1mm}3.98 & \hspace{-1mm}--    & \hspace{-1mm} --  \\
001328$+$135828 & \hspace{-1mm}3.576 & \hspace{-1mm}18.83 & \hspace{-1mm}$-0.69$ & \hspace{-1mm}6.68 & \hspace{-1mm}3.277 & \hspace{-1mm}14.79\\
001502$+$001212 & \hspace{-1mm}2.852 & \hspace{-1mm}18.79 & \hspace{-1mm}$-1.03$ & \hspace{-1mm}5.55 & \hspace{-1mm}--    & \hspace{-1mm} --  \\\hline
\end{tabular}
\end{center}
\end{table}

\section{DLA dust-reddening}\label{sec:dust}

\subsection{Spectral index distributions}\label{ssec:alpha}

\begin{figure}
\centerline{\includegraphics[height=80mm,angle=270]{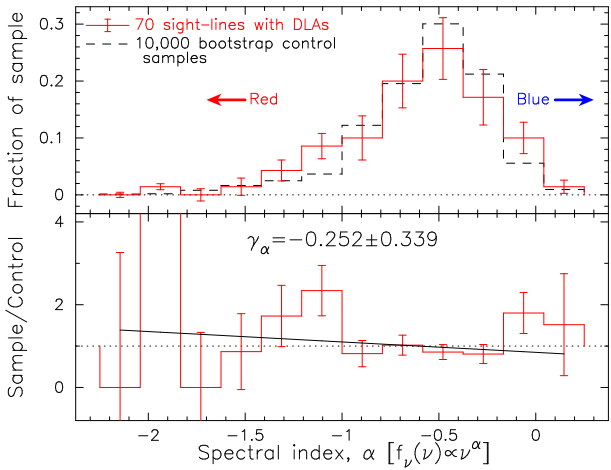}}
\caption{Top panel: Spectral index distributions of DLA and combined
  control samples. Error bars represent the rms over all bootstrap
  samples. Bottom panel: DLA-to-control number ratio with straight-line
  fit. The slope, $\gamma_\alpha$, is consistent with zero, i.e.~no
  evidence for DLA dust-reddening.}
\label{fig:alpha}
\end{figure}

The spectral index, $\alpha$, defined as $f_\nu\!\propto\!\nu^\alpha$, was
determined for each QSO by iteratively fitting a power-law to the flux
(corrected for Galactic extinction) $>\!10000\,\kms$ red-wards of the \lya\
emission line. Pixels $>\!4\,\sigma$ below or $>\!2\,\sigma$ above the fit
were rejected at each iteration and iterations continued until no more
points were rejected. Fig.~\ref{fig:example} shows an example fit. This
procedure effectively ignores intervening absorbers, intrinsic QSO
emission/absorption lines and the poorly sky-subtracted regions at
$\ga\!7000{\rm \,\AA}$ commonly seen in SDSS spectra. The statistical
errors in $\alpha$ range from $0.01$ to $0.35$ with $\ga\!80$\,per cent of
errors between $0.03$ and $0.15$.

\citet{AbazajianK_04a} discussed the DR2 spectrophotometric accuracy. There
is a $0.03{\rm \,mag}$ dispersion in the difference between the
$r^\prime-i^\prime$ fibre colours and those derived by convolving the
calibrated spectra with the filter transmission profiles. Thus, the error
on the mean colour difference between a sample of 70 sight-lines and a
large control sample is $\approx\!0.03/\!\sqrt{70}\!=\!0.004{\rm
\,mag}$. Hence, uncertainty in the spectrophotometry induces a mean
spectral index difference of just
$\left|\Delta\alpha\right|\!\approx\!0.01$ \citep{VandenBerkD_01a}.

To form a representative control sample to which the DLA sample may be
reliably compared, we drew 10000 bootstrap samples, each comprising 70
sight-lines, from the non-detection sample {\it with the same redshift
distribution as the DLA sample}. In practice, this was acheived by dividing
the DLA sample's range in \zem\ into 12 bins and randomly selecting from
the non-detection sample according to the relative populations in these 12
bins. There were 1160 non-detections over the \zem\ range of the DLA
sample. Fig.~\ref{fig:alpha} compares the $\alpha$ distributions of the DLA
and combined bootstrap sample. Both are very similar to the `photometric
spectral index' distribution for SDSS QSOs derived by
\citet{RichardsG_03a}. The error-bars represent the rms deviation in the
number of non-detections in each bin over the 10000 bootstrap samples.

To assess any differential reddening between the DLA and control samples,
the number of DLA sight-lines per bin is divided by the number in the
combined control sample for that bin, normalized by the total sample
sizes. The lower panel of Fig.~\ref{fig:alpha} shows this fraction fitted
by a straight line with slope $\gamma_\alpha\!=\!-0.25 \pm 0.34$ which
varies by $<\!0.4\,\sigma$ with different binning. The expected dispersion
in $\gamma_\alpha$ can be compared with the statistical error quoted here
by treating each bootstrap sample in the same way as the DLA sample,
deriving a slope $\gamma_{\rm boot}$ for each. The distribution of
$\gamma_{\rm boot}$ has rms $\sigma_{\rm boot}\!=\!0.36$ centred on
$\gamma_{\rm boot}\!=\!0.0$ and is well-fitted by a Gaussian.  That is, the
significance of any reddening is
$\left|-0.25\right|/0.36\,\sigma\!=\!0.69\,\sigma$, which has a Gaussian
probability of $P_{\rm G}\!=\!0.51$. This compares well with the
probability that the DLA and combined control samples are drawn from the
same parent distribution, $P_{\rm KS}\!=\!0.52$ and $P_{\rm MW}\!=\!0.49$,
using the Kolmogorov-Smirnov test and Mann-Whitney U-test respectively. All
three of the above statistical tests are most sensitive to differential
reddening in the bulk of the $\alpha$ distribution rather than at
$\alpha\!\la\!-1.3$ where the DLA and control samples contain fewer
sight-lines.

Fig.~\ref{fig:alpha} shows no evidence for dust-reddening of DR2 QSOs due
to foreground DLAs. By artificially altering the measured spectral indices
for the DLA sample by the same amount, $\delta\alpha$, we derived a simple
linear mapping, $\gamma_\alpha\!=\!-0.25+7.1\delta\alpha$. Using this
mapping, we derive a $1\,\sigma$ limit for the mean $\Delta\alpha$ allowed
by the data, $\Delta\alpha\!=\!-0.04 \pm 0.05$. This is inconsistent with
the claimed detection of dust-reddening by \citetalias{PeiY_91a},
$\Delta\alpha\!=\!-0.38 \pm 0.13$. Assuming a Small Magellanic Cloud (SMC)
extinction law for the DLAs and using the fitting formula of
\citet{PeiY_92a}, our limit on $\Delta\alpha$ corresponds to a $1\,\sigma$
limit on the mean $\EBV\!<\!0.007$. This does not exclude severe
dust-reddening in some small fraction of DLAs. Indeed, the control sample
suggests that the DLA with $\alpha\!\approx\!-1.9$ may be dust-reddened.

Another method for detecting DLA dust-reddening is to analyse the
$r^\prime\!-\!z^\prime$ distributions with a similar technique to that
above. Using the point spread function (PSF) magnitudes, we see no
differential colour: $\Delta(r^\prime\!-\!z^\prime)\!=\!0.025 \pm
0.026$. Our results also seem inconsistent with the $2\,\sigma$ detection
of dust-reddening found from the colour distributions of 2dF QSOs with and
without strong foreground Mg{\sc \,ii} absorbers \citep{OutramP_01a}. For
120 systems they find a mean colour excess $\EBV\!\approx\!0.04$ in the
observed frame. However, direct comparison with our results is difficult
since the Mg{\sc \,ii} selection means the absorbers have lower redshift
and may have significantly higher metallicities than our DLA sample.

The SDSS is a colour-selected QSO survey and so is biased against
intrinsically very red and heavily dust-reddened spectra. Therefore, our
results do not rule out a population of extremely dust-reddened QSOs
\citep[e.g.][]{GreggM_02a}. Applying a SMC extinction law with
$\EBV\!=\!0.1{\rm \,mag}$ (i.e.~$\Delta\alpha\!\approx\!-1.2$) to simulated
QSO spectra, \citet{RichardsG_03a} find the SDSS QSO survey completeness to
be largely unchanged. That is, the $\alpha$ distributions in
Fig.~\ref{fig:alpha} are sensitive to $\Delta\alpha\!\ga\!-1.0$ without
large biases due to colour-selection. \citeauthor{RichardsG_03a} also find
that only 6\,per cent of QSOs have spectra consistent with SMC-like
dust-reddening with $\EBV\!>\!0.04{\rm \,mag}$ and they interpret the
reddening to be internal to the QSO host-galaxy. This is consistent with
our results.

\begin{figure}
\centerline{\includegraphics[height=68mm,angle=270]{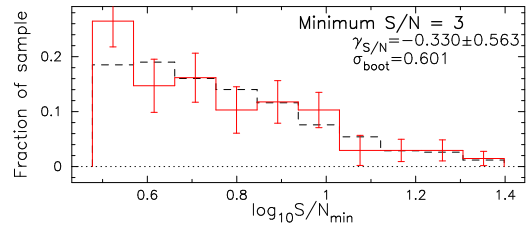}}
\caption{S/N distribution for the DLA (solid histogram) and combined
  control (dashed histogram) samples for our fiducial \snmin\
  threshold. Error bars are determined as in Fig.~\ref{fig:alpha}. At lower
  \snmin\ we may miss some low-S/N DLAs but $\snmin\!\geq\!3$ provides
  reliable DLA selection.}
\label{fig:SNR}
\end{figure}

\subsection{Robustness, potential biases and selection
  effects}\label{ssec:dbias}

To test the robustness of the above null result to possible biases and/or
selection effects, the data were subjected to the following tests, the
numerical results of which are summarized in Table \ref{tab:tests}.

{\bf Test 1: $\bmath{\snmin}$}. The \snmin\ threshold is a free parameter
in our DLA selection process.  If dust-reddening is significant and some
low-S/N DLAs are missed, the DLA sample may be biased against dust-reddened
QSOs. Fig.~\ref{fig:SNR} shows the distribution of \snmin\ for the DLA and
control samples analysed in a similar way to the spectral indices. There is
no evidence for a deficiency of low-S/N DLA detections, indicating that a
threshold of $\snmin\!=\!3$ is adequate. Table \ref{tab:tests} shows the
lack of dust-reddening to be robust against lower and higher \snmin\
thresholds, despite the former resulting in some relative incompleteness at
low \snmin.

\begin{table}
\begin{center}\vspace{-1mm}
\caption{Slopes and bootstrap rms (in parentheses) for different
  sub-samples and systematic error checks.  $N$ is the number of QSOs with
  foreground DLAs. Tests 1a, b \& c use \snmin\ thresholds of 2, 3 and 4
  respectively. All other tests use our fiducial value,
  $\snmin\!=\!3$. Tests 6a \& b are for the low- and high-$z$ samples and
  tests 7a \& b are for the low- and high-$W_{\rm r}$ samples
  respectively. See text for descriptions of other tests.}
\label{tab:tests}\vspace{-4mm}
\begin{tabular}{cccccc}\hline
\multicolumn{1}{c}{Test}&\multicolumn{1}{c}{\hspace{-1mm}$N$}&\multicolumn{1}{c}{\hspace{-1mm}\gama}&\multicolumn{1}{c}{\hspace{-1mm}\gamr}&\multicolumn{1}{c}{\hspace{-1mm}\gami}&\multicolumn{1}{c}{\hspace{-1mm}\gamz}\\\hline
1a      &\hspace{-1mm}81      &             \hspace{-1mm}$-$0.01(0.30) &             \hspace{-1mm}$-$0.65(0.17) &             \hspace{-1mm}$-$0.62(0.17) &             \hspace{-1mm}$-$0.65(0.17) \\
{\bf 1b}&\hspace{-1mm}{\bf 70}&\hspace{-1mm}$\bmath{-}${\bf 0.25(0.36)}&\hspace{-1mm}$\bmath{-}${\bf 0.48(0.20)}&\hspace{-1mm}$\bmath{-}${\bf 0.44(0.20)}&\hspace{-1mm}$\bmath{-}${\bf 0.51(0.19)}\\
1c      &\hspace{-1mm}50      &             \hspace{-1mm}$-$0.03(0.45) &             \hspace{-1mm}$-$0.69(0.26) &             \hspace{-1mm}$-$0.62(0.25) &             \hspace{-1mm}$-$0.67(0.25) \\
2       &\hspace{-1mm}50      &                \hspace{-1mm}0.21(0.44) &             \hspace{-1mm}$-$0.04(0.24) &             \hspace{-1mm}$-$0.05(0.24) &                \hspace{-1mm}0.00(0.23) \\
3       &\hspace{-1mm}66      &             \hspace{-1mm}$-$0.22(0.37) &             \hspace{-1mm}$-$0.45(0.21) &             \hspace{-1mm}$-$0.40(0.21) &             \hspace{-1mm}$-$0.46(0.20) \\
4       &\hspace{-1mm}64      &             \hspace{-1mm}$-$0.43(0.44) &             \hspace{-1mm}$-$0.42(0.21) &             \hspace{-1mm}$-$0.41(0.21) &             \hspace{-1mm}$-$0.50(0.21) \\
5       &\hspace{-1mm}29      &                \hspace{-1mm}0.06(0.50) &             \hspace{-1mm}$-$0.51(0.31) &             \hspace{-1mm}$-$0.42(0.30) &             \hspace{-1mm}$-$0.43(0.29) \\
6a      &\hspace{-1mm}37      &                \hspace{-1mm}0.34(0.50) &             \hspace{-1mm}$-$0.43(0.30) &             \hspace{-1mm}$-$0.34(0.28) &             \hspace{-1mm}$-$0.39(0.28) \\
6b      &\hspace{-1mm}33      &             \hspace{-1mm}$-$0.62(0.52) &             \hspace{-1mm}$-$0.43(0.28) &             \hspace{-1mm}$-$0.46(0.28) &             \hspace{-1mm}$-$0.55(0.28) \\
7a      &\hspace{-1mm}41      &             \hspace{-1mm}$-$0.48(0.49) &             \hspace{-1mm}$-$0.31(0.26) &             \hspace{-1mm}$-$0.28(0.25) &             \hspace{-1mm}$-$0.36(0.25) \\
7b      &\hspace{-1mm}29      &                \hspace{-1mm}0.22(0.52) &             \hspace{-1mm}$-$0.76(0.31) &             \hspace{-1mm}$-$0.72(0.31) &             \hspace{-1mm}$-$0.77(0.31) \\\hline
\end{tabular}
\end{center}
\end{table}

{\bf Test 2: GL}. In Section \ref{sec:grav} we assess the magnification of
QSOs due to foreground DLAs. If both the GL effect and DLA reddening are
significant then reddened QSOs will be preferentially brightened above our
\snmin\ threshold. We test this by artificially dimming the DLA sample with
a representative value for the GL magnification from Section
\ref{sec:grav}, $A=0.35{\rm \,mag}$: $A$ is added to $r^\prime$, $i^\prime$
and $z^\prime$ and \snmin\ is reduced accordingly for each DLA
sight-line. Table \ref{tab:tests} shows only a marginal increase in $\gama$
for this test.

\begin{figure*}
\hbox{
  \centerline{
    \includegraphics[height=56mm,angle=270]{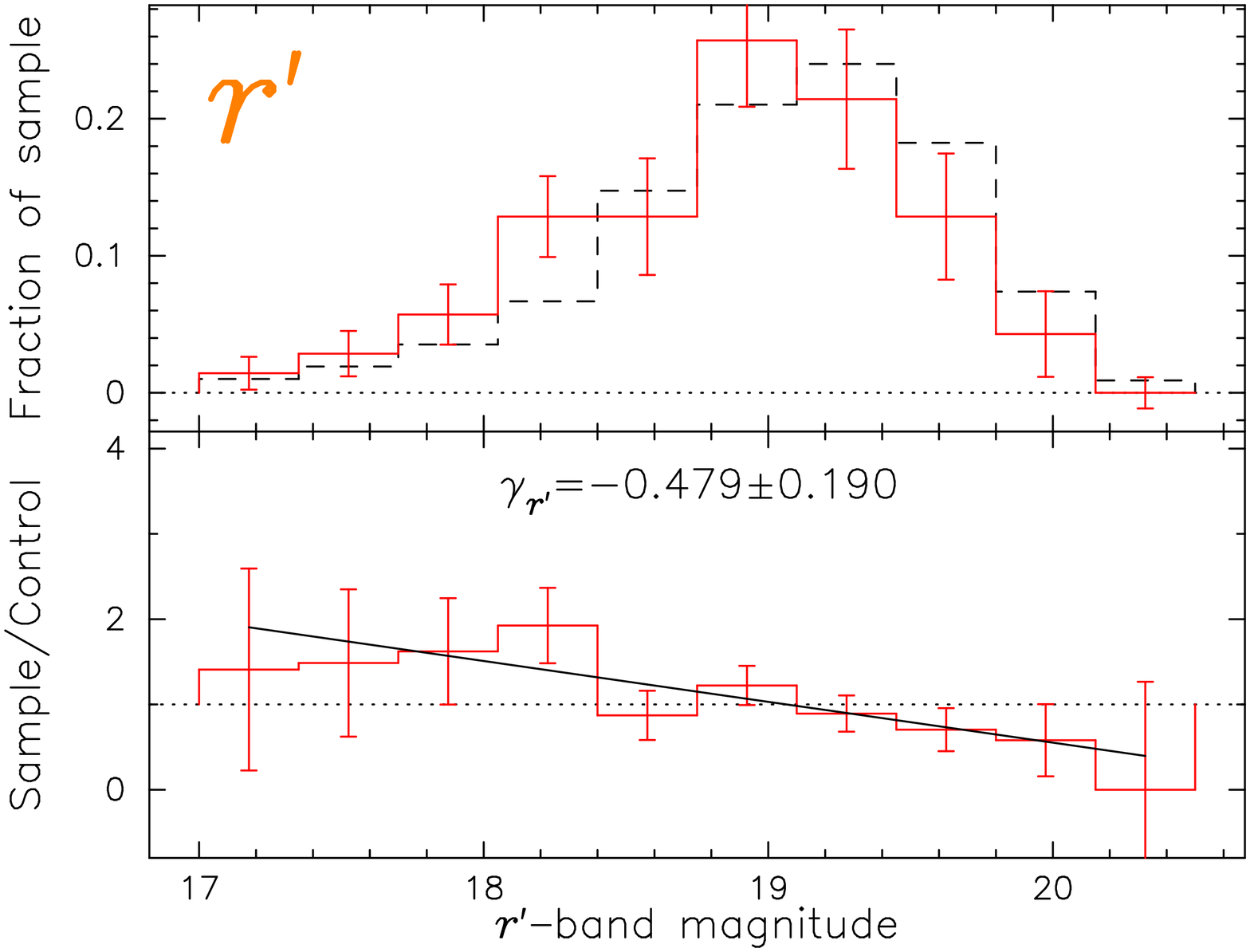}
    \hspace{1mm}
    \includegraphics[height=56mm,angle=270]{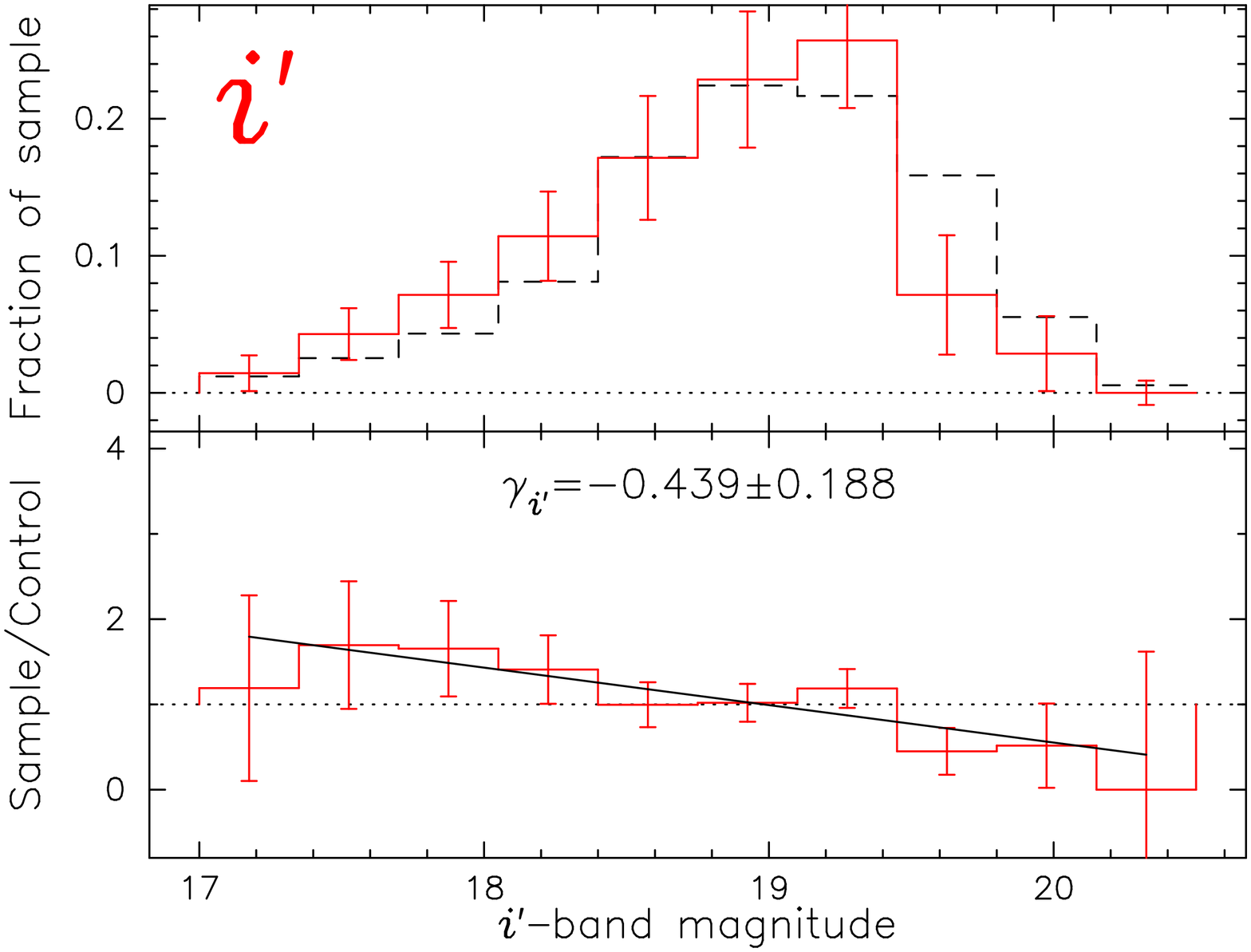}
    \hspace{1mm}
    \includegraphics[height=56mm,angle=270]{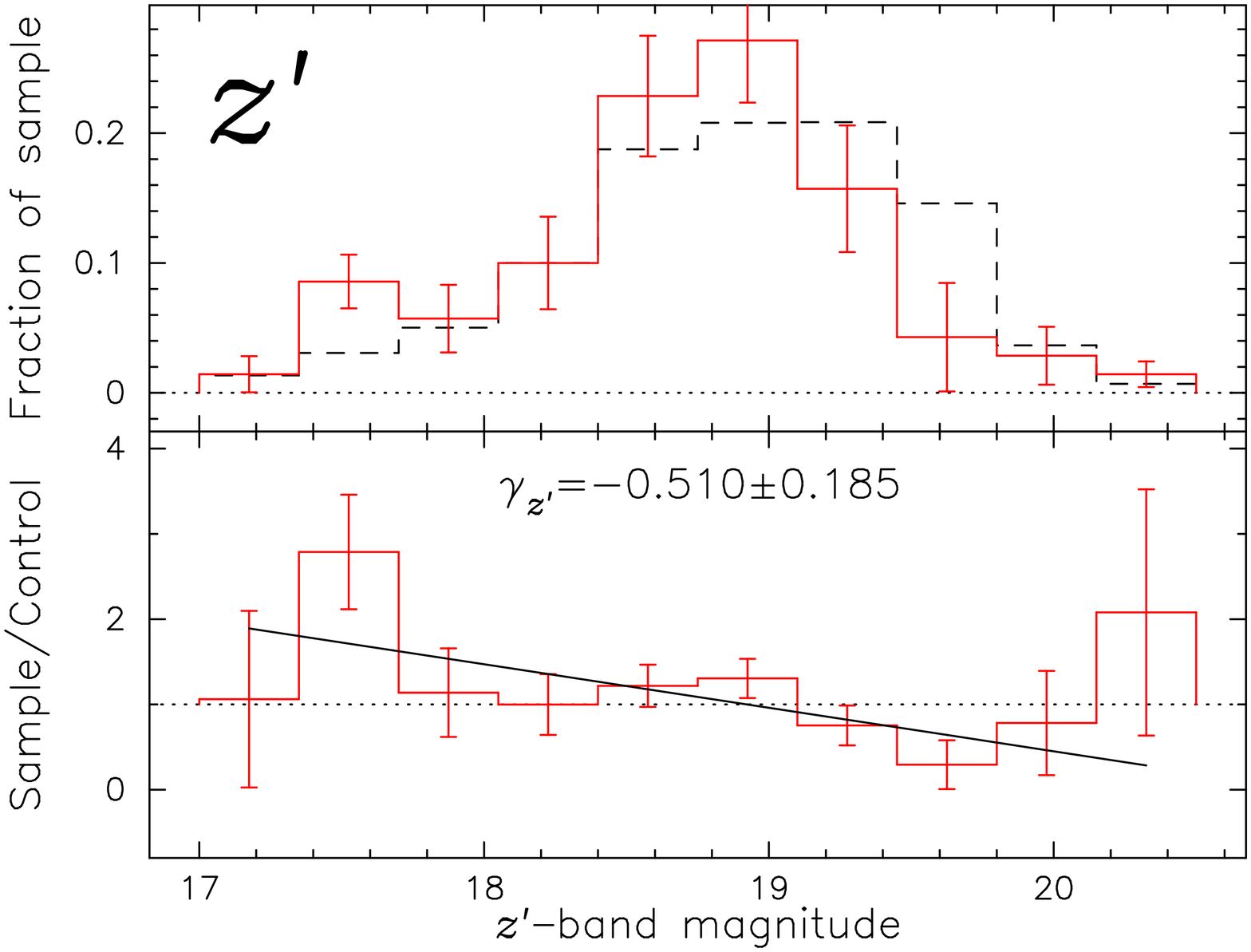}
  }
}
\caption{Top panels: PSF magnitude distributions for the DLA (solid histogram)
  and combined control (dashed histogram) samples. Error bars are
  determined as in Fig.~\ref{fig:alpha}. Bottom panels: DLA-to-control
  ratio with straight-line fit. The slopes, $\gamma_j$, indicate an excess
  of bright and/or a deficit of faint QSOs with DLAs.}
\label{fig:mags}
\end{figure*}

{\bf Test 3: Primary targets}. All SDSS QSOs were included in our sample,
rather than just those targeted by SDSS as QSOs based on their photometric
colours. This represents a slight inhomogeneity in our sample selection. We
have tested this by removing all `secondary' QSO targets from the sample
and repeating the analysis. The lack of DLA reddening is robust to this
test (Table \ref{tab:tests}).

{\bf Test 4: BALs}. Only the most severe BAL QSOs were removed from the
sample in Section \ref{sec:DLAs} since DLAs could still be easily detected
towards moderate BAL QSOs. However, it is possible (though improbable) that
heavily saturated Fe{\sc \,iii} $\lambda$1122 BALs may mimic DLAs. It is
also well known that BALs are somewhat redder than `normal' QSOs
\citep[e.g.][]{ReichardT_03a}, though any differential colour between DLA
and control samples should not be effected. To test these possibilities we
removed those QSOs which, by visual inspection, have some BAL features,
particularly near the C{\sc \,iv} emission line. A similar proportion
($\sim\!10$\,per cent) of such BALs was found in the DLA and control
samples. Once again, Table \ref{tab:tests} shows robust results.

{\bf Test 5: DR1}. \citet{SchneiderD_03a} formed a homogeneous QSO
catalogue from the SDSS DR1 \citep{AbazajianK_03a}. We applied our analysis
to this DR1 QSO sample using the DR2 spectra since \citet{AbazajianK_04a}
note the spectrophotometry of the DR2 is much improved. Although the DLA
sample is small, we find consistent results using the
\citeauthor{SchneiderD_03a} selection (Table \ref{tab:tests}).

{\bf Test 6: $\bmath{\zem}$-split}. The rest-frame composite SDSS QSO
spectrum of \citet{VandenBerkD_01a} shows an increased continuum level
between $2200$\,\AA\ and $4000$\,\AA\ due to many Fe{\sc \,ii} emission
lines. The SDSS spectra extend to $9200$\,\AA\, so our fitting method will
recover systematically redder spectral indices at $z_{\rm
em}\!\la\!3.2$. Again, there should be no differential shift between DLA
and control sample. Nevertheless, we split the samples at $z_{\rm
em}\!=\!3.2$ as a general consistency test. Table \ref{tab:tests} shows
the DLA sample to be somewhat bluer (redder) than the control sample for
the low-$z$ (high-$z$) portions. Though only marginally significant, we
note this shift here and will test it with larger data-sets in future work.

{\bf Test 7: $\bmath{\EWr}$-split}. The rest equivalent width of the DLA
line should be a measure of the H{\sc \,i} column density, $N$(H{\sc
\,i}). $\EWr({\rm Ly}\alpha)$ was determined over a rest-frame
$\Delta\lambda_{\rm r}\!=\!15{\rm \,\AA}$ window and so, for heavily damped
systems, some absorption may be missed. Continuum errors and Ly$\alpha$
forest blending also contribute significantly to errors in $\EWr({\rm
Ly}\alpha)$. Nevertheless, if high-$N$(H{\sc \,i}) DLAs cause more
dust-reddening, splitting the sample into low- and high-\EWr\
($>\!12.5$\,\AA) sub-samples may reveal this. Table \ref{tab:tests} shows
no significant difference between the $\alpha$ distributions for these
sub-samples.

To test the robustness of the result against the incompleteness at
low-$N($H{\sc \,i}$)$ mentioned in Section \ref{sec:DLAs}, we have also
analysed the $\alpha$ distributions for the DLA detections and
non-detections found by \citetalias{ProchaskaJ_04a}. Just 35 out of the 64
detected DLA sight-lines pass our (more restrictive) selection
criteria. The measured slope, $\gamma_\alpha\!=\!-0.91 \pm 0.62$,
corresponding to $\Delta\alpha\!=\!-0.13 \pm 0.09$, may indicate some
evidence for dust-reddening and, therefore, some evidence for high
dust-to-gas ratios in low-$N($H{\sc \,i}$)$ DLAs. This result is dominated
by just two quite red ($\alpha\!\approx\!-1.5$) DLA sightlines: removing
these two QSOs from the analysis gives $\gamma_\alpha\!=\!-0.42 \pm 0.64$
($\Delta\alpha\!=\!-0.06 \pm 0.09$). Interestingly, these two DLAs have
quite low $N($H{\sc \,i}$)$ according to PHF04, $\log_{10}N($H{\sc
\,i}$)\!\approx\!20.4$--$20.5$. Thus, a potentially important caveat to the
above non-detection of DLA dust-reddening is that, due to incompleteness of
the DLA selection at low-$N($H{\sc \,i}$)$, our method is somewhat
insensitive to an anti-correlation between dust-to-gas ratio and $N($H{\sc
\,i}$)$ in DLAs.

\section{DLA Gravitational lensing}\label{sec:grav}

\subsection{Magnitude distributions}\label{ssec:mags}

Following the analysis of \citetalias{MenardB_03a}, Fig.~\ref{fig:mags}
compares the DLA and control PSF magnitude distributions, corrected for
Galactic extinction. Fig.~\ref{fig:mags} shows an excess of bright and/or a
deficit of faint QSOs with DLAs relative to the control sample. The
best-fitting slopes, $\gamma_j$, to the DLA/control ratio and the rms of
$\gamma_j$ for the bootstrap samples are given in Table \ref{tab:tests}
(test 1b) for each band $j=r^\prime$, $i^\prime$ and $z^\prime$. All slopes
are significant at $>\!2\,\sigma$.

\begin{figure}
\centerline{\includegraphics[height=68mm,angle=270]{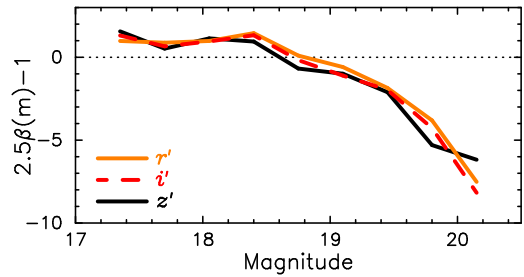}}
\caption{$2.5\,\beta(m)-1$ as a function of $r^\prime$, $i^\prime$ and
  $z^\prime$ PSF magnitude. GL will cause an excess (deficit) of bright
  (faint) QSOs with foreground DLAs.}
\label{fig:beta}
\end{figure}

Is the effect in Fig.~\ref{fig:mags} the signature expected from GL?
\citetalias{MenardB_03a} note two main competing effects: (i) the flux
density from the QSO is increased by the magnification factor $\mu$, and
(ii) the solid angle in which lensed QSOs appear is increased, reducing the
probability of observing them. \citetalias{MenardB_03a} show that a
relative excess or deficit of lensed QSOs is expected depending on $\mu$
and the gradient, $\beta$, of the control sample source counts as a
function of magnitude, $m$: $n(m)/n_0(m) \propto \mu^{2.5\,\beta(m)-1}$,
where $n/n_0$ is the number ratio of lensed to unlensed QSOs and
$\beta(m)\!\equiv\!{\rm d}\log_{10}[n_0(m)]/{\rm d}m$. Fig.~\ref{fig:beta}
shows $2.5\,\beta(m)-1$ for the $r^\prime$, $i^\prime$ and $z^\prime$
bands: given the control samples in Fig.~\ref{fig:mags}, GL should produce
a relative excess of lensed QSOs for $m\!\la\!19$ and a relative deficit
for $m\!\ga\!19$ in all bands. Note that the SDSS magnitude limit for
$\zem\!<\!3$ QSOs is $i^\prime\!=\!19.1$ whereas that for $\zem\!>\!3$ is
$i^\prime\!=\!20.2$ \citep{RichardsG_02a}. Effectively, this implies that
the derived faint-end slopes in Fig.~\ref{fig:beta} are too
negative. However, using only $\zem\!>\!3$ QSOs we see no appreciable
change in Fig.~\ref{fig:beta}. The results in Fig.~\ref{fig:mags} are
therefore {\it qualitatively} consistent with a GL interpretation.

As in \citetalias{MenardB_03a}, a simple illustrative example demonstrates
plausible {\it quantitative} agreement: consider a DLA at an impact
parameter of $10{\rm \,kpc}$ from a lens with an isothermal matter
distribution and velocity dispersion $\sigma_v\!=\!200\,\kms$. For
$\zem\!=\!3.3$, $\zab\!=\!2.8$, $\Omega_{\rm m}\!=\!0.3$,
$\Omega_\Lambda\!=\!0.7$ and $H_0\!=\!70\,\kms{\rm \,Mpc}^{-1}$, the
magnification factor is $\mu\!\approx\!1.06$. Fig.~\ref{fig:beta} therefore
implies expected gradients $\gamma_j\!\approx\!-0.15$ in
Fig.~\ref{fig:mags}. A more detailed comparison clearly requires precise
knowledge of the DLA impact parameters and host-galaxy halo-masses.

\subsection{Robustness and potential systematic effects}\label{ssec:gbias}

Despite the above result's low statistical significance ($\ga\!2\,\sigma$),
Table \ref{tab:tests} shows it is quite robust. Test 1 is particularly
important since, by setting the \snmin\ threshold too low, DLAs towards
fainter QSOs may be preferentially missed and the effect observed in
Fig.~\ref{fig:mags} may be artificially produced. \citet{EllisonS_04a} note
that such an effect may have produced a GL-like signature in their sample
of 47 strong Mg{\sc \,ii} absorbers at $0.6\!<\!\zab\!<\!1.7$. Indeed, with
a \snmin\ threshold of 2\,per pixel, where we do find a deficit of low-S/N
DLA-bearing sight-lines, Table \ref{tab:tests} shows more negative slopes,
$\gamma_j$. However, we see similar results even with a more conservative
threshold of 4\,per pixel where DLA detection is much more reliable.

\citetalias{MenardB_03a} explored some alternative explanations for the
putative GL effect. One important potential systematic error was DLA dust
obscuration producing a relative excess of faint QSOs in the DLA sample. In
Section \ref{sec:dust} we derived a $1\,\sigma$ limit on the colour excess
induced by SMC-like dust in the DLAs, $\EBV\!<\!0.007{\rm \,mag}$. This
corresponds to a total extinction of just $A_V\!\approx\!0.02{\rm \,mag}$
in the rest-frame $V\!$-band of the DLAs. Therefore, the GL magnification
dominates the dust obscuration in the SDSS DLA sample.

\section{Discussion and conclusions}

The SDSS provides a homogeneous QSO database which is ideal for studying
DLA properties with respect to large, carefully selected control
samples. We have utilised the SDSS DR2 to search for two important effects
DLAs may have on background QSO light: dust-reddening and gravitational
lensing (GL).

We find no evidence for dust-reddening of QSOs by foreground DLAs. The 70
sight-line DLA sample has a spectral index distribution consistent with
that of our combined control sample (Fig.~\ref{fig:alpha}), ruling out
overall shifts of $\left|\Delta\alpha\right|\!>\!0.19$ at $3\,\sigma$. This
corresponds to a limit on the colour excess due to SMC-like dust of
$\EBV\!<\!0.02{\rm \,mag}$ ($3\,\sigma$). Note that this is broadly
consistent with the reddening expected from the level of Fe depletion with
respect to Zn found in most DLAs. For a typical DLA with metallicity ${\rm
[Zn/H]}\!=\!-1.5$ and dust-depletion factor ${\rm [Zn/Fe]}\!=\!0.3$, the
dust-to-gas ratio is $\kappa\!\approx\!0.02$ times that found in the local
ISM. For $N($H{\sc \,i}$)\!=\!10^{21}{\rm \,cm}^{-2}$, this implies a shift
in $\alpha$ of just $\Delta\alpha\!\approx\!-0.06$ \citepalias{PeiY_91a} or
$\EBV\!\approx\!0.005{\rm \,mag}$ for a SMC extinction law. The concept of
a `typical' $\kappa$ for DLAs is, however, a poor one since [Zn/H] and
[Zn/Fe] vary from DLA to DLA by more than 1.5 and 0.8\,dex
respectively. For example, $\kappa\!\ga\!0.1$ is found for DLAs containing
H$_2$ \citep{LedouxC_03a}. Nevertheless, future comparison of
dust-reddening and depletion in a large sample of DLAs may lead to
constraints on DLA dust grain size and/or composition.

Our results are inconsistent with the those of \citetalias{PeiY_91a} who
found a mean $\Delta\alpha\!=\!-0.38 \pm 0.13$. The metallicities and
dust-depletion factors for \citetalias{PeiY_91a}'s DLAs are not especially
different to those of the overall DLA population. We suggest that
small-number statistics may have affected their results. Two notable
differences between the SDSS and \citetalias{PeiY_91a} QSO samples are that
the SDSS contains much fainter QSOs and extends to slightly higher \zem\
and \zab. Though an increase in $\kappa$ between the median \zab\ of the
two samples ($\zab\!\approx\!2.8$ and $\approx\!2.2$) may contribute to
this difference, evolution strong enough to entirely explain the
discrepancy is unlikely given the results of various abundance studies
\citep[e.g.][]{ProchaskaJ_03b,MurphyM_04b,CurranS_04c}. Since SDSS is a
colour-selected survey, a population of DLAs with high $N($H{\sc \,i}$)$
and high $\kappa$ cannot be ruled out, leaving open concerns about biases
in current estimates of the cosmological neutral gas mass density,
$\Omega_{\rm g}$ \citep[e.g.][]{BoisseP_98a}. However, our results provide
direct evidence that dust-extinction is quite low in the known DLA sample,
consistent with indirect estimates from abundance studies which have been
used to argue that any heavily reddened population of DLAs is small
\citep{ProchaskaJ_02a}. Finally, we note that our DLA selection is
incomplete at low-$N($H{\sc \,i}$)$ and so our result is relatively
insensitive to a possible (though seemingly unlikely) anti-correlation
between $\kappa$ and $N($H{\sc \,i}$)$ in DLAs.

We have also identified a possible signature of gravitational magnification
of QSOs by foreground DLAs in a similar vein as
\citetalias{MenardB_03a}. We expect an excess of bright QSOs with DLAs and
a deficit of faint QSOs with DLAs relative to our control sample, where the
dividing line should fall at $\sim\!19$th magnitude in $r^\prime$,
$i^\prime$ and $z^\prime$. This is indeed what is observed
(Fig.~\ref{fig:mags}). The amplitude of gravitational magnification,
measured from the slope of the DLA/control ratio versus magnitude, also
broadly agrees with that expected in a simple model of the QSO-DLA lensing
system. The putative lensing signal, though significant only at
$\ga\!2\,\sigma$, is robust against a variety of systematic error and
bias checks. Furthermore, a higher equivalent width sub-sample of DLAs
gives a stronger signal, as would be expected if these DLAs had lower
impact parameters. Refining the above results with future SDSS samples is
clearly important for future constraints on the dark matter halos of DLA
host-galaxies.

\section*{Acknowledgments}

We thank Bob Carswell, Michael Fall, Paul Hewett and Max Pettini for
discussions and the referee, Jason Prochaska, for a speedy and helpful
review. MTM thanks PPARC for support at the IoA under the observational
rolling grant. Funding for the creation and distribution of the SDSS
Archive has been provided by the Alfred P. Sloan Foundation, the
Participating Institutions, the National Aeronautics and Space
Administration, the National Science Foundation, the U.S. Department of
Energy, the Japanese Monbukagakusho, and the Max Planck Society.


\bsp_small

\label{lastpage}

\end{document}